\documentclass[]{emulateapj}
\usepackage{lscape}
\begin{document}
\title{Deep Mixing and Metallicity: Carbon Depletion in Globular Cluster Giants}
 
\bigskip
\author{Sarah L. Martell and Graeme H. Smith} 
\affil{University of California Observatories/Lick Observatory}
\affil{Department of Astronomy \& Astrophysics, UC Santa Cruz}
\affil{1156 High St, Santa Cruz, CA 95064}
\email{martell@ucolick.org, graeme@ucolick.org}
\and 
\author{Michael M. Briley}
\affil{Department of Physics and Astronomy, University of Wisconsin Oshkosh}
\affil{800 Algoma Blvd, Oshkosh, WI 54901}
\email{mike@maxwell.phys.uwosh.edu}
\bigskip
\begin{abstract} 
\noindent

We present the results of an observational study of the efficiency of deep mixing in globular cluster red giants as a function of stellar metallicity. We determine [C/Fe] abundances based on low-resolution spectra taken with the Kast spectrograph on the 3m Shane telescope at Lick Observatory. Spectra centered on the $4300\mbox{\AA}$ CH absorption band were taken for 42 bright red giants in 11 Galactic globular clusters ranging in metallicity from M92 ([Fe/H]$=-2.29$) to NGC 6712 ([Fe/H]$=-1.01$). Carbon abundances were derived by comparing values of the CH bandstrength index $S_{2}(CH)$ measured from the data with values measured from a large grid of SSG synthetic spectra. Present-day abundances are combined with theoretical calculations of the time since the onset of mixing, which is also a function of stellar metallicity, to calculate the carbon depletion rate across our metallicity range. We find that the carbon depletion rate is twice as high at a metallicity of [Fe/H]$=-2.3$ than at [Fe/H]$=-1.3$, which is a result qualitatively predicted by some theoretical explanations of the deep mixing process.

\end{abstract}


\section{Introduction}
The observation that carbon abundance in globular cluster red giants declines 
continuously as the stars evolve has inspired a great deal of observational and
theoretical study (e.g., Suntzeff 1981\nocite{S81}, Carbon et al. 
1982\nocite{C82}, Trefzger et al. 1983\nocite{T83}, Suntzeff \& Smith 
1991\nocite{SS91}, Weiss \& Charbonnel 2004\nocite{WC04}, Denissenkov \& Tout 2000\nocite{DT00}, Smith \& Briley 2006\nocite{SB06}, and similar work). 
Canonically, abundances should be static  on the red giant branch after the 
first dredge-up because of the broad radiative zone between the 
hydrogen-burning shell and the surface. Progressive carbon depletion with 
rising luminosity on the giant branch is commonly interpreted as a sign of a 
non-convective ``deep mixing'' process that mixes carbon-depleted material from
the hydrogen-burning shell, where the CN(O) cycle is acting, to the surface 
(e.g., Sweigart \& Mengel 1979\nocite{SM79}, Charbonnel 1995\nocite{C95}, 
Charbonnel et al. 1998\nocite{CBW98}, Bellman et al. 2001\nocite{BBSC01}, 
Denissenkov \& VandenBerg 2003\nocite{DV03}, and others). This same depletion 
of surface carbon abundance is also observed in red giants in the halo field 
(e.g., Gratton et al. 2000\nocite{GSCB00}), and is observed to occur at the 
same rate in the halo field as in globular clusters with halo-like
 metallicities (e.g., Smith \& Martell 2003\nocite{SM03}). 

The process of deep mixing, inferred from observations of low [C/Fe], 
$\log \epsilon$(Li), and $^{12}$C/$^{13}$C, is only observed to occur in stars 
brighter than the red giant branch (RGB) luminosity function ``bump'' 
\citep{CBW98}. There are indications (e.g., Langer et al. 1986\nocite{L86}, Bellman et al. 2001\nocite{BBSC01}) that there may be carbon depletion in 
stars fainter than the RGB bump in the metal-poor globular cluster M92, 
though there is not an obvious physical explanation for such a 
phenomenon. During the first dredge-up, in the subgiant phase, the base of the 
convective envelope drops inward to smaller radius as the stellar core 
contracts, mixing the partially-processed material of the stellar interior with
the unprocessed material at the surface. As hydrogen shell burning progresses, 
low on the giant branch, the temperature gradient in the star steepens and the 
base of the convective envelope begins to move outward, leaving behind a sharp 
jump in mean molecular weight (the ``$\mu$-barrier'') at the point of its 
furthest inward progress \citep{I65}. As the convective envelope retreats 
outward, this steep $\mu$ gradient finds itself within a radiative region 
between the hydrogen-burning shell and the base of the convective zone, where 
it can potentially hinder mass motions within the radiative zone. 

The red giant branch bump is an evolutionary stutter that occurs when the 
hydrogen-burning shell, which is advancing outward in mass, encounters the 
$\mu$ barrier. The sudden influx of hydrogen-rich material to the 
hydrogen-burning shell causes the star to become briefly bluer and fainter 
before it re-equilibrates and continues to evolve along the red giant branch 
(e.g., Iben 1968\nocite{I68}, Cassisi et al. 2002\nocite{C02}). In a collection
of stars with equal age and composition, this evolutionary loop will result in 
an unexpectedly large number of stars at a particular magnitude, and a bump in 
the differential luminosity function. At a fixed mass, the base of the 
convective envelope sinks lower in higher-metallicity stars during the first 
dredge-up, meaning that the RGB bump occurs at a fainter luminosity on the RGB 
in high-metallicity globular clusters than in low-metallicity clusters (e.g., 
Zoccali 1999\nocite{Z99}). However, because evolutionary timescales shorten as 
stellar mass rises, there is a maximum mass of $\simeq 2$M$_{\odot}$ for stars to
experience this evolutionary loop: above that mass, the hydrogen-burning shell 
does not move outward far enough to cross the $\mu$ barrier in the short time 
the star is on the RGB \citep{G89}. The fact that deep mixing does not begin 
until after the hydrogen-burning shell crosses the $\mu$ barrier is interpreted
by, e.g., \citet{C94} to mean that the gradient of mean molecular weight is the
dominant factor in permitting or prohibiting deep mixing. Indeed, \citet{DV03} 
point out that $\nabla \mu$ is ``the only physical quantity that changes 
significantly while approaching the hydrogen-burning shell'' in post-bump red 
giants. \citet{CPT05} provide a somewhat different perspective: in their maximal-mixing models, the $\mu$ gradient inhibits mixing on the upper giant branch, but rotational mixing processes are not strong enough on the lower giant branch to cause observable changes in surface abundances, regardless of whether there is a steep $\mu$ gradient present.

The underlying physical reason for deep mixing is not clear, though rotation 
has commonly been implicated since \citet{SM79} proposed meridional circulation
as an explanation for CNO anomalies in red giants. Recent theoretical studies 
tend to focus on specific parametrizations and representations of the process; 
for example, Rayleigh-Taylor instability \citep{EDL08}, diffusion \citep{DV03},
and thermohaline mixing \citep{CZ07}. \citet{P06} demonstrated that meridional 
circulation, differential rotation, and shear turbulence do not create enough 
mixing to account for the observed variations in surface abundances, implying 
that additional hydrodynamical processes must be acting. The large study of 
surface abundances in field giants published by \citet{GSCB00} is a 
key to differentiating between models of deep mixing, since it demonstrates 
clearly the progressive depletion of carbon on the giant branch, as well as the
sharp drop in $^{12}$C/$^{13}$C and $\log \epsilon$(Li) that happens at the RGB
bump.

The fundamental result from \citet{GSCB00} that all current deep mixing models 
must reproduce is that deep mixing is universal among post-bump red giants. 
\citet{CBW98} consider mixing in terms of the ``critical $\mu$ gradient,'' the 
largest gradient in mean molecular weight that still permits deep mixing. In an
observational study of seven mildly metal-poor red giants in the region of the 
RGB bump, they find that the critical $\mu$ gradient is independent of 
composition or mass.

\citet{DV03} use the formalism of diffusion, with mixing depth and a diffusion 
constant as the important parameters, to model deep mixing. They find that the 
mixing depth does not depend strongly on metallicity, which implies that all 
red giants with a mass less than $\simeq 2$M$_{\odot}$ will experience deep 
mixing. Their figures also show that the evolution of surface abundances of 
carbon and nitrogen are not particularly affected by metallicity, though 
reductions in $\log \epsilon$(Li) and $^{12}$C/$^{13}$C are more sensitive. 
\citet{EDL08} show that the reaction $^{3}$He($^{3}$He, 2p)$^{4}$He causes a 
$\mu$-inversion in the outer edge of the hydrogen-burning shell, and claim that
the resulting Rayleigh-Taylor instability is important in driving deep mixing. 
\citet{CZ07} argue that the more complex process of thermohaline convection 
will act in that $\mu$-inversion region. They use the \citet{U72} prescription 
to parametrize the thermohaline mixing as a diffusion process. In contrast to 
\citet{DV03}, they find that the evolution of the surface abundances of carbon,
nitrogen, and lithium are all affected by overall stellar metallicity, while 
the $^{12}$C/$^{13}$C ratio approaches its equilibrium value very quickly at 
all metallicities.

Although questions of deep mixing rate (e.g., Smith \& Martell 
2003\nocite{SM03}) and depth (e.g., Charbonnel et al. 1998\nocite{CBW98}) have 
been studied observationally by many authors, the results available in the 
literature can be difficult to synthesize into a single conclusion. Many 
studies focus on one or two particular clusters (e.g., Da Costa \& Cottrell 
1980\nocite{DC80}, Suntzeff 1981\nocite{S81}, Trefzger et al. 1983\nocite{T83}, Lee 1999\nocite{L99}), or attempt to correlate deep mixing with other 
cluster properties such as horizontal branch morphology \citep{CN00}, stellar 
rotational velocity \citep{CPT05} or cluster ellipticity \citep{N87}. 
Individual authors and collaborations develop their own analysis tools, and the
differences between spectral index definitions, model atmospheres, spectral 
synthesis engines, and abundance determination methods produce significant 
systematic differences in different authors' abundance scales, as is clear from
literature-compilation studies such as \citet{S02}.

One can construct a phenomenological picture of deep mixing from this 
heterogeneous information, and it goes roughly as follows: all stars with mass 
less than $\simeq 2$M$_{\odot}$ will at some point have their hydrogen-burning 
shell cross the $\mu$-barrier. The $\mu$-barrier is larger than the critical 
$\mu$-gradient for deep mixing, so its destruction permits deep mixing to 
begin. Deep mixing occurs continuously, and involves all material outside the 
radius where the $\mu$-gradient within the outer H-burning shell is 
critical. The onset of deep mixing happens lower on the giant branch for 
higher-metallicity clusters, because their $\mu$-barrier is at smaller radius. However, 
in higher-metallicity stars the hydrogen-burning shell is more 
compact \citep{SM79}, so that the radius where the $\mu$-gradient is 
critical is relatively further out in the hydrogen-burning shell. This means that the material mixed to the surface in higher-metallicity stars is less processed than in low-metallicity stars. Various authors (e.g., Charbonnel et al. 1998\nocite{CBW98}, Cassisi et al. 2002\nocite{C02}) use 
this relation between metallicity and mixing efficiency to study the structure 
of the hydrogen-burning shell.

Our goal in this project is to determine the relative efficiency of deep mixing
across a broad range of metallicity by measuring present-day carbon abundances 
and depletion rates from a homogeneous set of globular cluster red 
giants in similar evolutionary phases. An earlier example of this approach is 
the study of \citet{BD80}, who found that [C/Fe] on the upper RGB of 
M3, M13 and NGC 6752 correlated with [Fe/H] metallicity.

\section{The Data Set}
In order to construct a data set that could be used to compare CH bandstrengths and [C/Fe] abundances across a wide range of metallicity, we chose to obtain our data exclusively with the Kast double spectrograph on the Shane 3m telescope at Lick Observatory, devoting 21 nights between July 2004 and August 2006 to the data collection. Using a mirror in place of a dichroic, we directed all light to the blue side of the spectrograph, where the 600/4310 (moderate-resolution) grism produced a pixel spacing of $\simeq 1.8\mbox{\AA}/$pix and a resolution of approximately $5.4\mbox{\AA}$ over a wavelength range of 3400 to 5400$\mbox{\AA}$. The detector at the time was a thinned 1200 $\times$ 400-pixel Reticon CCD. Table 1 lists names, positions, distance moduli, reddenings, metallicities, and number of stars observed for each globular cluster included in the survey. Metallicities, distance moduli and reddenings were all taken from the February 2003 revision of the online compilation of \citet{H96}. Photometry for the individual globular clusters was taken from a combination of original photographic color-magnitude diagram work and proper-motion membership studies: \citet{SW55} for NGC 4147, \citet{S53} and \citet{C79_M3} for M3, \citet{A55} and \citet{C79_M5} for M5, \citet{A55} and \citet{C79_M13} for M13, \citet{A55} and \citet{HRdR76} for M10, \citet{C76_M92} for M92, \citet{C76_M15} for M15, \citet{L80} for NGC 6535, \citet{SS66} and \citet{C88_6712} for NGC 6712, \citet{B65} for M56, and \citet{H75} and \citet{CR87_M2} for M2.

\begin{figure}[]
\includegraphics[width=\columnwidth]{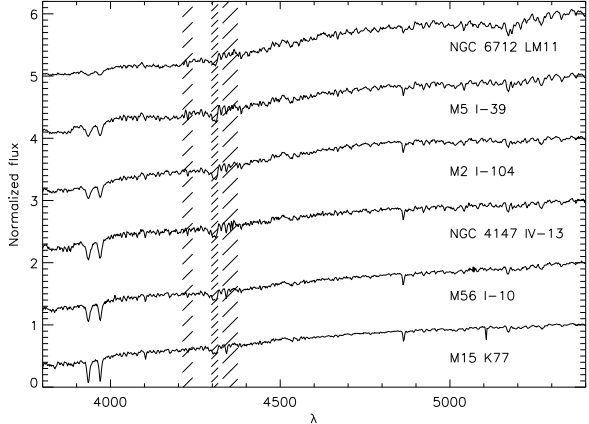} 
\caption{
Selected spectra from our data set. $M_{V}\simeq -1.5$ for all stars, [Fe/H] varies from $-2.25$ (M15 K77) to $-1.01$ (NGC 6712 LM11) in roughly equal steps. The vertical bars with widely spaced shading mark the bandpasses for the comparison bands of $S_{2}(CH)$, and the narrowly spaced shading marks the corresponding science band.
\label{fig:LBf1}}
\end{figure}

Targets were observed with a slit width of $1\arcsec$, $1.5\arcsec$, or $2\arcsec$ depending on the seeing. Ideally, three consecutive exposures of each star were taken, with typical exposure times of 1800s, to allow for cosmic ray removal. For 17 of the 42 stars in our sample, there was only time to obtain one or two exposures on a particular night: four of them were observed again at a later date, nine have only two exposures, and four have only one exposure. In all cases, the various exposures were reduced independently, and spectra of each individual star were coadded to produce a final (unfluxed) 1-d spectrum. The standard stars Feige 34, Feige 110, BD +28 4211, and BD +33 2642 were observed with a $9\arcsec$ slit aligned with the parallactic angle in an effort to capture maximal UV flux for reliable flux calibrations. Since Kast sits at Cassegrain focus, spectra of the HeHgCd lamp were taken directly following target spectra, with the same telescope pointing, slit width and dispersive element, to facilitate wavelength calibration and account for flexure.

Data reduction was accomplished with the XIDL LowRedux package made available by J.X. Prochaska at UC Santa Cruz\footnote{available from http://www.ucolick.org/\char126xavier/LowRedux/}. This comprehensive, updated version of the XIDL code used in earlier form in \citet{MSB08} handles bias subtraction, flat-fielding, cosmic ray removal, object identification and extraction, sky subtraction, flexure correction, wavelength calibration, atmospheric correction, coadding and flux calibration. 

Our targets are bright red giants in the range $-1.0 \geq M_{V} \geq -2.0$, which generally required 3600-second exposures with Kast to obtain signal-to-noise ratios per pixel just redward of the G band of roughly 150. These stars are all significantly brighter than the RGB luminosity function bump, meaning that any reasonable deep mixing rate will have had time to make a measurable decrement in surface carbon abundance. Figure \ref{fig:LBf1} shows selected spectra from our sample, with metallicities ranging from [Fe/H]$=-2.29$ (lowest spectrum) to [Fe/H]$=-1.01$ (highest spectrum) in roughly equal steps. As [Fe/H] rises, there is a clear increase in the strength of the Mg b and MgH features near $5170\mbox{\AA}$, a slight increase in the strength of the broad CN absorption feature at $4215\mbox{\AA}$, and a reddening of the overall continuum shape that reduces the apparent depth of the Ca II H\& K lines at 3935 and 3970$\mbox{\AA}$ and the CN bandhead at 3883$\mbox{\AA}$.

\section{Analysis}
We use the index $S_{2}(CH)$, recently defined in \citet{MSB08b} to be sensitive to carbon abundance and relatively independent of nitrogen abundance over a wide range of metallicity, to quantify the strength of the CH G band in all of our combined, flux-calibrated spectra. As with most spectroscopic indices, $S_{2}(CH)$ is measured as the magnitude difference between the integrated flux in the relevant absorption feature (the ``science band'') and the integrated flux in two nearby relatively absorption-free bands (the ``comparison bands''). As discussed in \citet{MSB08b}, the science band for $S_{2}(CH)$ runs from 4297$\mbox{\AA}$ to 4317$\mbox{\AA}$, and the comparison bands run from 4212$\mbox{\AA}$ to 4242$\mbox{\AA}$ and 4330$\mbox{\AA}$ to 4375$\mbox{\AA}$. The bandpasses for $S_{2}(CH)$ are shown as shaded regions in Figure \ref{fig:LBf1}, with the more widely spaced lines marking the comparison bands, and the more closely spaced shading lines marking the science band.

One-sigma errors on measured values of $S_{2}(CH)$ were determined as in \citet{MSB08}: for stars observed three or more times, $\sigma_{S}$ is calculated as the standard deviation on the mean of the individual index values measured from flux-calibrated, uncombined spectra, and for stars observed twice, $\sigma_{S}=\frac{0.89 \times \Delta S_{2}(CH)}{\sqrt{2}}$. Stars observed only once are assumed to have errors in their $S_{2}(CH)$ values equal to the mean value of $\sigma_{S}$, which is 0.0047.

\begin{figure}[]
\includegraphics[width=\columnwidth]{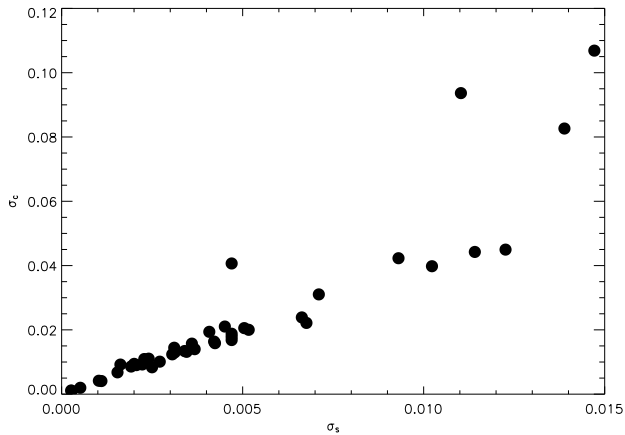} 
\caption{
Error in calculated [C/Fe] abundance resulting from error in the measured index $S_{2}(CH)$, for each star in our sample. The abundance error $\sigma_{\rm[C/Fe]}$ is never larger than 0.12 dex, and is almost always smaller than 0.05 dex.
\label{fig:LBf2}}
\end{figure}

To convert bandstrengths to [C/Fe] abundances, we follow a method similar to that described in \citet{MSB08}, matching measured values of $S_{2}(CH)$ to values derived from synthetic spectra interpolated to match the data in $M_{V}$ and [Fe/H] (the ``model grid''). We select the subset of the model grid with a canonical [N/Fe] value of $+0.6$ and interpolate between model $S_{2}(CH)$ values to find a preliminary [C/Fe]. We then interpolate the model grid to that preliminary [C/Fe] value, allow [N/Fe] to have its full range of possible values, and interpolate between model values of the CN bandstrength index $S(3839)$ \citep{N81} to calculate a preliminary [N/Fe] abundance. If that preliminary [N/Fe] is more than 0.1 dex lower or higher than the canonical value, we interpolate the model grid to match it and repeat the $S_{2}(CH)$-matching process to obtain a final [C/Fe] abundance. As a check, we also re-calculate [N/Fe] using the final [C/Fe] value, and we find that the difference between preliminary and final [C/Fe] and [N/Fe] is never large. This is due in large part to the nitrogen-insensitivity of $S_{2}(CH)$: for a fixed [C/Fe], [Fe/H] and $T_{\rm eff}$, varying [N/Fe] by large amounts does not change $S_{2}(CH)$ significantly.

The synthetic spectra employed are quite similar to those used in \citet{MSB08}: values for $T_{\rm eff}$ and $\log g$ were taken from 12-Gyr Yale-Yonsei \citep{D04} isochrones calculated for each cluster metallicity, and MARCS model atmospheres \citep{G75} and the SSG spectral synthesis program (Bell, Paltoglou, \& Trippico 1994 and references therein\nocite{BPT94}) were used to generate synthetic spectra. For each individual globular cluster metallicity, [C/Fe] varies from $-1.4$ to $+0.4$ in steps of 0.2 dex, and [N/Fe] varies from $-0.6$ to $+2.0$ in steps of 0.2 dex. As in \citet{MSB08}, other variables such as $^{12}{\rm C}/^{13}{\rm C}$, [O/Fe], and $v_{turb}$ were chosen to be consistent with the values used in \citet {BC01}. Synthetic spectra were smoothed to a resolution of 5.4$\mbox{\AA}$ and a pixel spacing of 1.8$\mbox{\AA}$ to match the data.

\begin{figure}[]
\includegraphics[width=\columnwidth]{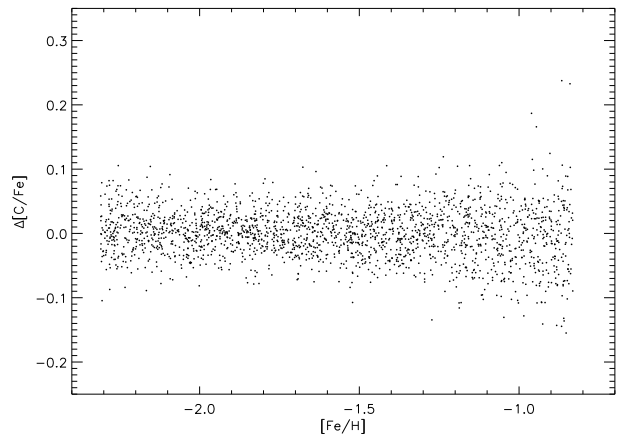} 
\caption{
Method-based error $\Delta$[C/Fe] as a function of [Fe/H], for synthetic spectra with $-2.31 \leq$[Fe/H]$\leq -0.83$, $-1.0 \leq$[C/Fe]$\leq 0.0$, $0.0 \leq$[N/Fe]$\leq 1.0$, and 2\% Poisson noise. Method-based errors for individual stars in our sample were assigned as the RMS of all points in this figure within $\pm 0.05$ of the [Fe/H] abundance of the parent cluster. Figure adapted from \citet{MSB08b}.
\label{fig:LBf3}}
\end{figure}

The errors on our [C/Fe] determinations have several sources, some of which are
readily quantifiable: the errors in the measured values of $S_{2}(CH)$ are 
generally quite small, and propagate into small noise-based [C/Fe] errors. We 
determine these noise-based [C/Fe] errors for each star in our sample as half 
the difference between [C/Fe] calculated for $S_{2}(CH)+\sigma_{S}$ and [C/Fe] 
calculated for $S_{2}(CH)-\sigma_{S}$. As can be seen in Figure \ref{fig:LBf2},
these errors are generally smaller than 0.05 dex, and are all smaller than 0.12
dex. They are also comparable in magnitude to the errors introduced by our 
index-matching method (the ``method-based error''). We calculate a
method-based error by choosing a random ([Fe/H], [C/Fe], [N/Fe]) point within 
the abundance range spanned by our model grid, interpolating to find the 
$S_{2}(CH)$ value that corresponds, and using that value to re-calculate [C/Fe]
assuming a canonical [N/Fe] value of $+0.6$. The difference between the 
original input value of [C/Fe] and the calculated value is the method-based 
error. It is discussed more thoroughly in \citet{MSB08b}, and is a complex 
function of all of the input abundances. Figure \ref{fig:LBf3}, which is 
adapted from \citet{MSB08b}, shows model-based error for a set of synthetic 
spectra with 2\% Poisson-distributed noise and abundances in the range 
typically inhabited by globular cluster stars ($-2.31 \leq$[Fe/H]$\leq -0.83$, 
$-1.0 \leq$[C/Fe]$\leq 0.0$, $0.0 \leq$[N/Fe]$\leq 1.0$), as a function of 
[Fe/H]. The addition of noise to the synthetic spectra was done to make them 
resemble the observed spectra they are compared to, so that the calculated 
method-based error would be an accurate representation of the actual error 
introduced by our carbon-determination method. We assign a 
method-based error for a given star in our survey as the RMS of all points in 
Figure \ref{fig:LBf3} that fall within $\pm0.05$ dex of the [Fe/H] metallicity 
of its parent cluster. 

We combine the noise-based error and the method-based error on calculated [C/Fe] in quadrature, since they are independent, and find that the combined error $\sigma_{C}$ never exceeds 0.12 dex, and is usually under 0.05 dex. Table 2 lists cluster name, star name, date observed, cluster [Fe/H], $M_{V}$, $S_{2}(CH)$, $\sigma_{S}$, [C/Fe], and $\sigma_{C}$ for each star in our sample. It should be noted that we do not include possible errors in our adopted values of [Fe/H] or the corresponding $T_{\rm eff}$ and $\log (g)$ values taken from the \citet{D04} isochrones in $\sigma_{C}$.

\begin{figure}[]
\includegraphics[width=\columnwidth]{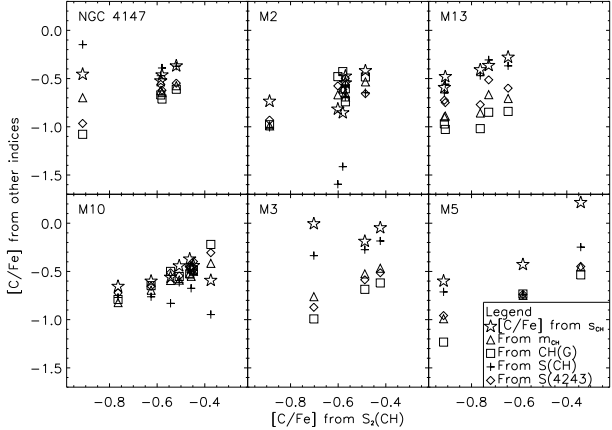} 
\caption{Comparison of [C/Fe] values calculated from various CH bandstrength indices, for six of the eleven clusters in our sample. There is generally a vertical offset between abundances derived from the various indices, which may be indicative of errors in flux calibration.
\label{fig:LBf4}}
\end{figure}

The matter of systematic offsets to our abundance scale resulting from our flux calibration, temperature scale, or choice of model atmosphere is not as straightforward to assess. We compiled Figure \ref{fig:LBf4} to explore the relationship between [C/Fe] values calculated by using a number of different G-band indices defined in the literature. Each vertical column of points represents one star, with [C/Fe] calculated from the indices $s_{CH}$ \citep{BS93}, $m_{CH}$ \citep{S96}, and $CH(G)$ \citep{L99} (which were created for studies of red giants in the moderate-metallicity globular clusters M13, M3, and M5) plotted as stars, triangles, and squares, respectively. Crosses and diamonds, respectively, represent [C/Fe] calculated from the indices $S(CH)$ \citep{MSB08} and $S(4243)$ \citep{B90}, which were designed to be used with red giants in the low-metallicity globular clusters M53, M55 and NGC 6397. The comparison and science bandpasses of these indices are given in Table 3. To determine these alternate [C/Fe] values, we followed an index-matching process identical to the one we used to calculate [C/Fe] from $S_{2}(CH)$.

Interestingly, in Figure \ref{fig:LBf4}, there tend to be consistent vertical offsets between [C/Fe] values from the various indices. Since each index covers slightly different wavelength regions, we interpret this to mean that each one captures different information from the spectrum. In general, indices tuned for low-metallicity stars can have wider comparison bands, and do not need to be as carefully defined as indices for high-metallicity clusters, because there are absorption features that are important to avoid in high-metallicity spectra (for example, the CN banhdead at 4215$\mbox{\AA}$). This makes the broader lower-metallicity indices less affected by noise in the observed spectra, but unreliable and often nitrogen-sensitive when used on high-metallicity spectra.

\begin{figure}[]
\includegraphics[width=\columnwidth]{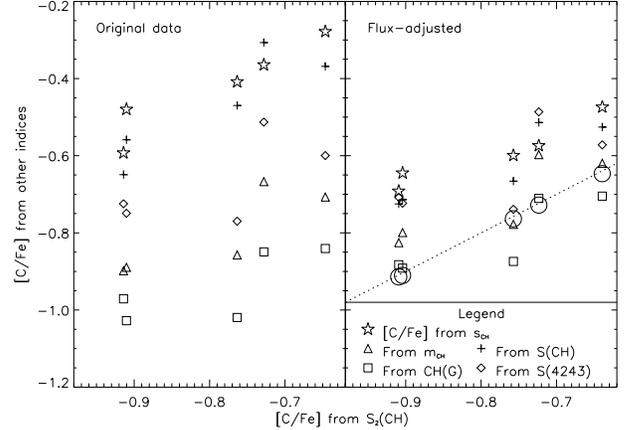} 
\caption{Results of the flux-adjustment test: vertical scatter among [C/Fe] values calculated for M13 stars from various CH-sensitive indices is greatly reduced when the overall shape of the data continuum is matched to the appropriate synthetic spectrum. The left panel shows [C/Fe] calculated from five G-band indices defined in the literature (y-axis) versus [C/Fe] calculated from $S_{2}(CH)$ (x-axis) for the five M13 stars in our data set. The right panel shows the results of the same measurements made on data adjusted to match the overall shape of abundance-matched synthetic spectra. Large circles in the right panel show the [C/Fe] values calculated from $S_{2}(CH)$ and the non-adjusted data spectra, and lie very close to the dashed line showing 1:1 correspondence.
\label{fig:LBf5}}
\end{figure}

In Figure \ref{fig:LBf4}, the values for [C/Fe] calculated from $s_{CH}$ are almost always the highest value for any particular star. Since $s_{CH}$ is defined to have only one comparison band, it is the most susceptible of all the indices we measure to errors in flux calibration or temperature scale. To investigate the possible effects of a mismatch between the overall shape of the data and the synthetic spectra, we performed a test on our M13 spectra, which have a fairly large range in [C/Fe] calculated from the various indices. For each of the five M13 stars in our sample, we adjusted the flux calibration to match the overall spectral shape of an appropriate synthetic spectrum. We fit a parabola to fairly absorption-free regions between 4000$\mbox{\AA}$ and 4500$\mbox{\AA}$ in both the data and interpolated-synthetic spectra, and used the ratio of those two curves to adjust the overall shape of the data to match the synthetic spectra. The left panel of Figure \ref{fig:LBf5} is the same as the M13 panel of Figure \ref{fig:LBf4}, and shows the [C/Fe] abundances calculated from the various CH indices. The right panel shows [C/Fe] abundances calculated from the same CH indices, measured from the flux-adjusted spectra, and the vertical scatter is considerably decreased. The large circles in the right panel show the $S_{2}(CH)$-derived [C/Fe] values, measured from the original (non-flux-adjusted) spectra, and they fall almost exactly on the dashed line showing 1:1 correspondence. We make two conclusions from this exercise: first, that disagreement between [C/Fe] calculated from indices that purport to measure the same underlying abundance can be a sign of flux-calibration or temperature-scale errors, and second, that continuum-division is likely preferable to flux-calibration, so that this potential source of error can be avoided altogether.

However, the [C/Fe] values calculated from $S_{2}(CH)$ are barely changed by the flux adjustment. As discussed in \citet{MSB08b}, $S_{2}(CH)$ was designed specifically to be fairly sensitive to carbon abundance and insensitive to nitrogen abundance over a wide range in metallicity. Therefore we feel that our [C/Fe] values are fairly robust, and we choose to use $S_{2}(CH)$ for carbon determinations in all of our data, rather than building a patchwork of metallicity-tuned indices and correcting for estimated systemic offsets. In addition, since our goal is a differential measurement of [C/Fe] between stars of varying [Fe/H], the absolute zeropoint of our [C/Fe] abundance scale is not vital to our result.

\begin{figure}[]
\includegraphics[width=\columnwidth]{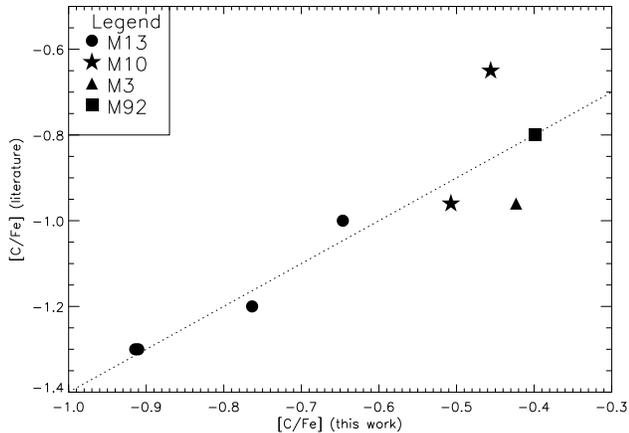} 
\caption{Comparison of our [C/Fe] values calculated from $S_{2}(CH)$ with literature carbon abundances. Literature data for M3 and M13 are from \citet{S81}, the M10 data are from \citet{SBH05}, and the M92 data point is taken from \citet{BBSC01}. The dashed line shows the line [C/Fe]$_{\rm this work}=$[C/Fe]$_{\rm lit}+0.4$, which is consistent with the literature-compilation studies of \citet{SB06} and \citet{S02}.
\label{fig:LBf6}}
\end{figure}

In Figure \ref{fig:LBf6} we plot [C/Fe] derived from our $S_{2}(CH)$ 
measurements versus [C/Fe] values taken from the literature, for the eight stars in
our sample with published carbon abundances. Carbon abundances in M3 and M13 were 
taken from \citet{S81}, carbon abundance values for M10 are taken from 
\citet{SBH05}, and the single point for M92 is from \citet{BBSC01}. The data 
compilations of \citet{SB06} and \citet{S02} find that \citet{S81} carbon 
abundances tend to fall $\simeq 0.35$ dex below other reported carbon 
abundances in M13 and M3. It is therefore encouraging to find that the 
dashed line in Figure \ref{fig:LBf6}, which shows the relation 
[C/Fe]$_{\rm lit} = $[C/Fe]$_{\rm this work} - 0.4$, traces the data so closely. As for
the M10 and M92 stars, our [C/Fe] values are clearly larger than those reported
in the literature, but with such small crossover samples it is difficult to 
describe this decisively as a systematic offset.

\section{Results}
A functional form for the evolution of carbon abundance with time may be found 
by considering the stellar envelope as a simple system with a constant mass 
$M$ and mixing rate $\dot{M}$. If deep mixing removes carbon from the 
combined convective envelope plus atmosphere at a rate $-\dot{M}\times X_{C, env}$ (where $X_{C, env}$ is the mass fraction of carbon in the envelope at time $t$), and introduces it at a rate 
$\dot{M}\times X_{C, hbs}$ (where $X_{C, hbs}$ is the mass fraction of carbon 
in the hydrogen-burning shell), then the rate of change of atmospheric carbon 
abundance can be written as 
\begin{eqnarray*}
\frac{d(MX_{C, env})}{dt}=\dot{M}(X_{C, hbs}-X_{C, env})
\end{eqnarray*}
The equilibrium abundance of carbon in the CNO cycle is quite small, so this 
integrates simply to 
\begin{eqnarray*}
X_{C, env}\propto \exp^{-(\dot{M}/M)t}
\end{eqnarray*}
where $M/\dot{M}$ can be thought of as the characteristic timescale for mixing.
Since [C/Fe] is a logarithmic measure of carbon abundance, an exponential 
decline in $X_{C, env}$ will result in a linear decline in [C/Fe] with time. 
\citet{SM03} find a linear decline in [C/Fe] with $M_{V}$ among bright red 
giants in the field and in the globular clusters M92, NGC 6397, and M3, with a 
slope of 0.22 dex/magnitude. The field-star study of \citet{GSCB00} finds a 
linear trend in [C/Fe] with $\log(L/L_{\odot})$. According to \citet{D04} 
evolutionary tracks, the $\log(L/L_{\odot})$-$t$ relationship is not exactly 
linear for post-bump RGB stars ($1.8 \leq \log(L/L_{\odot}) \leq 2.6$), but it 
is sufficiently close that a linear [C/Fe] - $M_{V}$ or [C/Fe] - $\log(L/L_{\odot})$ relation suggests a fairly linear [C/Fe] - $t$ relation as well.

\begin{figure}[]
\includegraphics[width=\columnwidth]{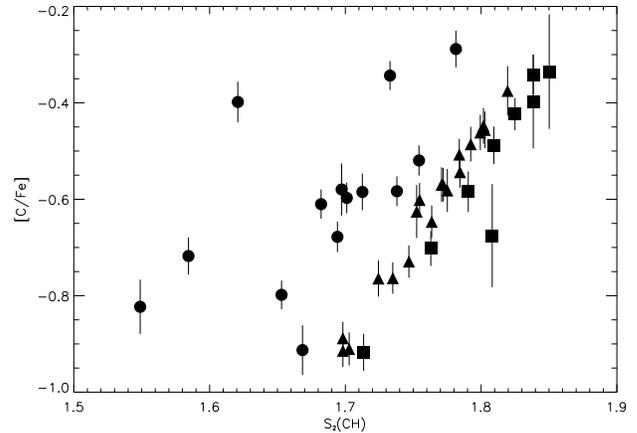} 
\caption{
Derived [C/Fe] abundance versus measured $S_{2}(CH)$ bandstrength for all stars in our survey, with total error bars as described in the text and symbols denoting three broad [Fe/H] bins. Circles represent stars with [Fe/H]$\leq -1.7$, triangles are for the bin $-1.7 \leq$[Fe/H]$\leq -1.4$, and squares show stars with [Fe/H]$\geq -1.4$. Equal [C/Fe] abundances are expressed as different CH bandstrengths depending on [Fe/H].
\label{fig:LBf7}}
\end{figure}

\begin{figure}[]
\includegraphics[width=\columnwidth]{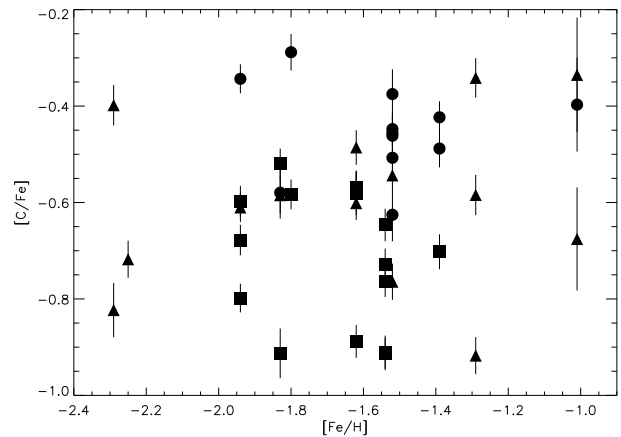} 
\caption{
Derived [C/Fe] abundance with error bars, as in Figure \ref{fig:LBf7}, versus [Fe/H]. Here symbols denote three broad $M_{V}$ bins. Circles show stars with $M_{V} \geq -1.3$, triangles are for the bin $-1.3 \geq M_{V} \geq -1.6$, and squares are for stars with $M_{V}\leq -1.6$. Although the faintest and brightest bins separate fairly well into high and low carbon abundances, respectively, it is unclear from this plot whether low-metallicity stars experience more rapid carbon depletion than high-metallicity stars. 
\label{fig:LBf8}}
\end{figure}

Figure \ref{fig:LBf7} shows calculated [C/Fe] versus measured $S_{2}(CH)$ for 
our sample, with symbols indicating three broad metallicity bins. Circles are 
stars from clusters with [Fe/H]$\leq -1.7$, triangles represent stars with 
$-1.7 \leq$[Fe/H]$\leq -1.4$, and squares are for stars with [Fe/H]$\geq -1.4$.
It can be clearly seen that equal carbon abundances are expressed as lower CH 
bandstrengths at lower metallicity. The error bars shown are equal to 
$\pm \sigma_{C}$. Figure \ref{fig:LBf8} shows calculated [C/Fe] with error bars
as a function of [Fe/H], and here the symbols represent three broad luminosity 
bins. These bins have the same sense as in Figure \ref{fig:LBf7}, with the 
faintest stars represented by circles and the brightest stars plotted as 
squares. Although the faintest and brightest stars occupy mostly separate 
regions in this figure, there is not a clear trend between [C/Fe] and [Fe/H]. However, the lack of a clear correlation in Figure \ref{fig:LBf8} does not necessarily mean that there is no dependence of deep mixing rate on stellar metallicity.

The question of how much time each observed star has spent in the deep mixing phase 
is the key to interpreting our data: we selected targets with similar absolute 
magnitudes, but it is well-known observationally \citep{FP90} and theoretically
\citep{CS97} that the RGB bump occurs at lower luminosity in more metal-rich 
globular clusters. In addition, low-metallicity red giants evolve more quickly 
than high-metallicity red giants of the same mass (e.g., Riello et al. 
2003\nocite{R03}). As a result, the $M_{V}=-1.5$ red giants in the metal-rich 
end of our sample have been mixing for far longer than their metal-poor 
counterparts. In order to study deep mixing efficiency with our data set, we 
must convert our [C/Fe] values, which are a function of both [Fe/H] and time, 
to some time-invariant quantity. We choose to do this by converting present-day
[C/Fe] abundances to a carbon depletion rate per Gyr. This is a two-step 
process: we must first calculate the absolute magnitude of the RGB 
bump, $M_{V}^{bump}$, for each individual cluster, and then for each star we must convert the $V$ magnitude height above the bump, $\Delta M_{V}^{bump}$, into a time since the onset of deep mixing. 

\begin{figure}[]
\includegraphics[width=\columnwidth]{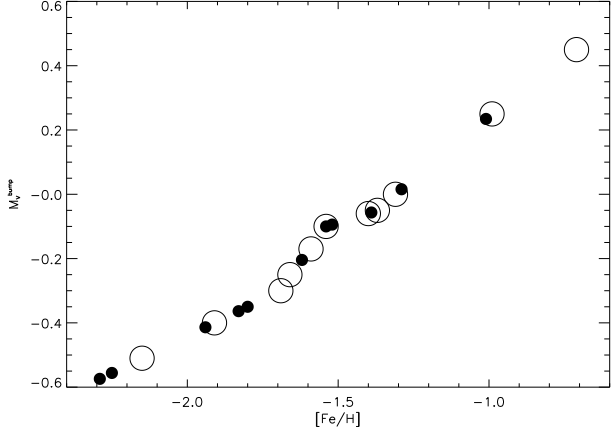} 
\caption{
The relation between the absolute $M_{V}$ magnitude of the ``bump'' in the red giant branch luminosity function and metallicity. Large circles are data from 11 globular clusters observed in \citet{FP90}, and small circles are the values for the globular clusters in our sample, interpolated linearly from the \citet{FP90} data.
\label{fig:LBf9}}
\end{figure}

\begin{figure}[]
\includegraphics[width=\columnwidth]{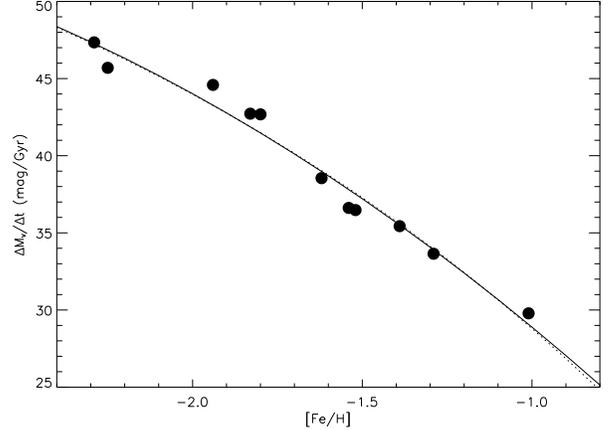} 
\caption{
Rate of evolution $\Delta M_{V}/\Delta t$ versus [Fe/H] for the eleven globular clusters in our sample, calculated using Yale-Yonsei \citep{D04} isochrones and evolutionary tracks in the range $M_{V}^{bump} \ge M_{V} \ge -1.5$. The solid curve is the best-fit polynomial, and the dashed curve is the best-fit exponential.
\label{fig:LBf10}}
\end{figure}

We use the observational study of \citet{FP90} to calculate $M_{V}^{bump}$ for 
each cluster in our survey. As can be seen in Figure \ref{fig:LBf9}, there is a
nearly-linear relationship between observed $M_{V}^{bump}$ (plotted as large 
open circles) and [Fe/H]. We calculate $M_{V}^{bump}$ for the eleven clusters 
in our sample by a simple linear interpolation from the \citet{FP90} data. The 
conversion from $\Delta M_{V}^{bump}$ to $\Delta t^{bump}$ is based on 
Yale-Yonsei \citep{D04} isochrones and evolutionary tracks. For each cluster 
metallicity, we created an evolutionary track for a star with appropriate 
metallicity and a mass taken from the $M_{V}=-1.5$ point in a 
metallicity-matched 12 Gyr YY isochrone. We then chose points in the 
evolutionary track near $M_{V}^{bump}$ and $M_{V}=-1.5$, and we calculate 
$\Delta M_{V}/\Delta t$ based on those points. Figure \ref{fig:LBf10} shows our
calculated values for $\Delta M_{V}/\Delta t$ for all clusters in our sample, 
along with a best-fit polynomial (solid curve) and a best-fit exponential 
(dashed curve), which are nearly identical. 

\begin{figure}[]
\includegraphics[width=\columnwidth]{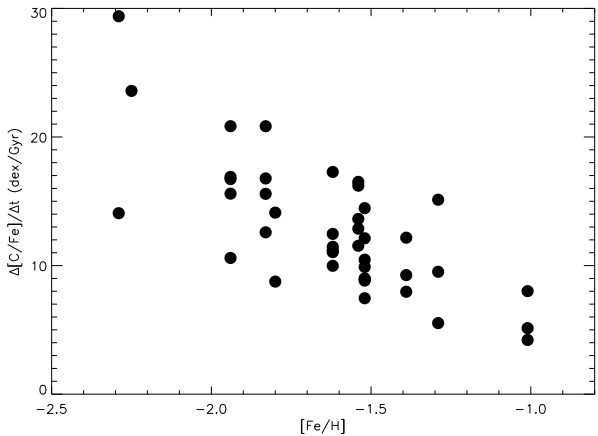} 
\caption{
Carbon depletion rate $\Delta$[C/Fe]$/\Delta t$ versus [Fe/H] for all of the stars in our sample. There is a clear downward trend with rising metallicity, a result predicted by the deep mixing models of, e.g., \citet{CZ07}.
\label{fig:LBf11}}
\end{figure}

Combining these two steps, we calculate the carbon depletion rate 
$\Delta$[C/Fe]$/\Delta t$ as [C/Fe]$/\Delta M_{V}^{bump} \times \Delta M_{V}/\Delta t$, assuming that the initial [C/Fe] abundance is solar, and that deep mixing begins at the RGB bump for all stars. Figure \ref{fig:LBf11} shows $\Delta$[C/Fe]$/\Delta t$ versus [Fe/H] for all stars in our sample, and there is a clear downward trend in carbon depletion rate as metallicity increases. We see no break or corner in the relation between $\Delta$[C/Fe]$/\Delta t$ and [Fe/H], indicating that mixing is never completely prohibited in our sample. This may happen at metallicities above [Fe/H]$ = -1.0$, if the hydrogen-burning shell is so compressed that the critical $\mu$-gradient is reached outside it. Our result is in agreement with the theoretical predictions of \citet{CZ07}, and confirms that the process of deep mixing is less effective in relatively high-metallicity red giants.

\section{Conclusions and Future Work}
In summary, the results of this work show that within the [Fe/H] range $-2.2$ to $-1.0$ dex the rate of deep mixing varies with metallicity. We measure CH bandstrength using the index $S_{2}(CH)$, which was designed to be valid across a broad range in [Fe/H] \citep{MSB08b}. Carbon abundances are determined by matching CH bandstrengths measured from the data to bandstrengths measured from specifically-designed grids of SSG synthetic spectra. Under the assumption \citep{CBW98} that deep mixing begins at the RGB luminosity function bump, we establish the carbon depletion rate for a given star as the change in its [C/Fe] from an initial solar value divided by the time elapsed since the onset of mixing. Since the RGB luminosity function bump is fainter in high-metallicity globular clusters than in lower-metallicity globular clusters, and higher-metallicity red giants evolve more slowly than lower metallicity red giants, higher-metallicity stars at a given absolute magnitude qualitatively ought to have spent more time mixing than their lower-metallicity counterparts. In order to quantify the time spent mixing as a function of metallicity and absolute V magnitude, we interpolate the \citet{FP90} observational data on $M_{V}^{bump}$ as a function of metallicity to the metallicities of the clusters in our study, then use metallicity-appropriate Yale-Yonsei \citep{D04} isochrones to convert $\Delta M_{V}^{bump}$ for each individual star into $\Delta t^{bump}$. As can be seen in Figure \ref{fig:LBf10}, the lower-metallicity red giants do evolve more quickly, and therefore spend less time mixing than the higher-metallicity red giants in this study. However, dividing [C/Fe] by $\Delta t^{bump}$ for each individual star produces the result that the deep mixing rate is roughly twice as large at [Fe/H]$=-2.0$ than at [Fe/H]$=-1.0$, as can be seen in Figure \ref{fig:LBf11}. 

This present analysis includes some assumptions worth noting, since they may 
need to be more carefully examined if this result is to provide constraints for
theoretical models of deep mixing. To express $X_{C, env}$ as a straightforward
exponential in time, we hold the mass of the stellar envelope constant, though 
it shrinks continuously as the hydrogen-burning shell proceeds outward 
\citep{DV03}. We also hold the mixing rate $\dot{M}$ constant, though it is 
controlled by the structure of the hydrogen-burning shell, and may well evolve.
In addition, we implicitly assume in constructing Figure \ref{fig:LBf11} that 
all stars have equal (and solar) initial abundances of carbon. A primordial 
depletion of $\sim 0.3$ dex in a subset of our sample would thus be 
misinterpreted as an artificially high mixing rate in those stars. A study of carbon abundances in fainter post-bump red giants, or pre-bump giants, in the globular clusters included in this study, would allow for direct calculation of $\Delta$[C/Fe]$/\Delta t$ without assumptions about the initial carbon abundance. In addition, a study of carbon depletion rates in high-metallicity red giants, either in old open clusters or in high-metallicity globular clusters, would allow an estimation of the maximum metallicity at which deep mixing still operates. That maximum metallicity would be helpful for constraining the various models of deep mixing. Nevertheless, our fundamental result, that carbon depletion proceeds more quickly in low-metallicity globular cluster red giants than in their high-metallicity counterparts, is robust, and provides a clear affirmation of present theoretical models.

\clearpage
\begin{deluxetable}{ l r r c c r c }
\tablewidth{0pt}
\tablehead{\colhead{Cluster ID} & 
\colhead{RA (J2000)} & 
\colhead{$\delta$ (J2000)} &
\colhead{$(m-M)_{V}$} &
\colhead{$E(B-V)$} & 
\colhead{[Fe/H]} & 
\colhead{$\rm{N_{obs}}$}}
\startdata
NGC 4147&12 10 06.2&+18 32 31&16.48&0.02&-1.83&4\\
NGC 5727 (M3)&13 42 11.2&-28 22 32&15.08&0.01&-1.39&3\\
NGC 5904 (M5)&15 18 33.8&+02 04 58&14.46&0.03&-1.29&3\\
NGC 6205 (M13)&16 41 41.5&+36 27 37&14.45&0.02&-1.54&5\\
NGC 6254 (M10)&16 57 08.9&-04 05 58&14.08&0.28&-1.52&8\\
NGC 6341 (M92)&17 17 07.3&+43 08 11&14.64&0.02&-2.29&2\\
NGC 6535&18 03 50.7&-00 17 49&15.22&0.34&-1.80&2\\
NGC 6712&18 53 04.3&-08 42 22&15.60&0.45&-1.01&3\\
NGC 6779 (M56)&19 16 35.5&+30 11 05&15.65&0.20&-1.94&5\\
NGC 7078 (M15)&21 29 58.3&+12 10 01&15.23&0.10&-2.25&1\\
NGC 7089 (M2)&21 33 29.3&-00 49 23&15.49&0.06&-1.62&6\\
\enddata
\end{deluxetable}

\begin{deluxetable}{ l l l r r r r r r }
\tablewidth{0pt}
\tablehead{\colhead{Cluster ID} &
\colhead{Star ID} &
\colhead{Date obs.} &
\colhead{$M_{V}$} &
\colhead{[Fe/H]} &
\colhead{$S_{2}(CH)$} &
\colhead{$\sigma_{S}$} &
\colhead{[C/Fe]} &
\colhead{$\sigma_{C}$}}
\startdata
M10&II-105&2006-06-03&-1.23&-1.52&1.753&0.0114&-0.625&0.0549\\
M10&III-73&2006-06-03&-1.22&-1.52&1.784&0.0011&-0.507&0.0327\\
M10&III-85&2006-06-03&-1.38&-1.52&1.784&0.5153&-0.544&0.0325\\
M10&IV-30&2006-06-03&-1.28&-1.52&1.803&0.0052&-0.456&0.0381\\
M10&I-15&2004-07-12&-1.22&-1.52&1.799&0.0047&-0.461&0.0365\\
M10&III-93&2004-07-12&-1.18&-1.52&1.802&0.0047&-0.448&0.0370\\
M10&III-97&2004-07-12&-1.41&-1.52&1.724&0.0047&-0.764&0.0375\\
M10&I-12&2005-07-12&-1.14&-1.52&1.819&0.0102&0.375&0.0513\\
M13&IV-53&2005-04-17&-1.77&-1.54&1.747&0.0019&-0.729&0.0336\\
M13&II-33&2006-06-01&-1.78&-1.54&1.764&0.0021&-0.647&0.0337\\
M13&III-52&2006-06-01&-1.78&-1.54&1.735&0.2635&-0.763&0.0325\\
M13&II-57&2006-06-02&-1.75&-1.54&1.703&0.0023&-0.910&0.0342\\
M13&III-18&2006-06-02&-1.68&-1.54&1.698&0.0020&-0.914&0.0338\\
M15&K77&2005-09-06&-1.53&-2.25&1.584&0.0016&-0.718&0.0386\\
M2&I-103&2005-07-13&-1.92&-1.62&1.775&0.0071&-0.582&0.0447\\
M2&I-104&2005-07-13&-1.52&-1.62&1.792&0.0042&-0.486&0.0359\\
M2&I-298&2005-09-07&-1.46&-1.62&1.755&0.0037&-0.601&0.0351\\
M2&I-190&2005-09-08&-1.75&-1.62&1.698&0.0024&-0.888&0.0340\\
M2&II-60&2005-09-09&-1.69&-1.62&1.772&0.0034&-0.570&0.0348\\
M2&II-71&2005-09-09&-1.64&-1.62&1.771&0.0042&-0.570&0.0361\\
M3&BC&2006-05-31&-1.24&-1.39&1.825&0.0010&-0.423&0.0333\\
M3&I-46&2006-05-31&-1.26&-1.39&1.809&0.0050&-0.488&0.0389\\
M3&V-80&2006-06-02&-1.61&-1.39&1.763&0.0031&-0.702&0.0361\\
M5&I-39&2006-06-01&-1.39&-1.29&1.838&0.0041&-0.342&0.0410\\
M5&IV-34&2006-06-02&-1.41&-1.29&1.791&0.0045&-0.584&0.0418\\
M5&IV-49&2006-06-03&-1.32&-1.29&1.713&0.0031&-0.917&0.0384\\
M56&I-10&2006-08-30&-1.53&-1.94&1.682&0.0022&-0.610&0.0304\\
M56&E-48&2006-08-31&-1.95&-1.94&1.694&0.0031&-0.678&0.0315\\
M56&I-141&2005-09-06&-1.22&-1.94&1.733&0.0025&-0.343&0.0301\\
M56&E-22&2005-09-07&-1.78&-1.94&1.701&0.0035&-0.597&0.0318\\
M56&I-66&2005-09-07&-1.78&-1.94&1.653&0.0015&-0.798&0.0297\\
M92&II-70&2006-05-31&-1.40&-2.29&1.621&0.0036&-0.398&0.0420\\
M92&IV-94&2006-05-31&-1.44&-2.29&1.549&0.0047&-0.823&0.0563\\
NGC 4147&II-14&2005-02-01&-1.150&-1.83&1.697&0.0123&-0.580&0.0539\\
NGC 4147&II-30&2005-02-01&-1.760&-1.83&1.754&0.0027&-0.519&0.0313\\
NGC 4147&II-45&2005-02-02&-1.960&-1.83&1.668&0.0093&-0.913&0.0516\\
NGC 4147&IV-13&2005-02-02&-1.390&-1.83&1.713&0.0066&-0.585&0.0381\\
NGC 6535&13&2004-07-13&-1.71&-1.80&1.738&0.2743&-0.583&0.0308\\
NGC 6535&19&2004-07-13&-0.990&-1.80&1.781&0.0068&-0.288&0.0379\\
NGC 6712&KC564&2005-07-13&-1.430&-1.01&1.850&0.0147&-0.335&0.1187\\
NGC 6712&LM11&2005-07-13&-1.200&-1.01&1.838&0.0139&-0.397&0.0974\\
NGC 6712&B66&2006-08-31&-1.57&-1.01&1.808&0.0110&-0.676&0.1069\\
\enddata
\end{deluxetable}

\clearpage
\begin{deluxetable}{l c c c l}
\tablewidth{0pt}
\tablehead{\colhead{Index} &
\colhead{Blue comparison band (\mbox{\AA})} &
\colhead{Science band (\mbox{\AA})} &
\colhead{Red comparison band (\mbox{\AA})} &
\colhead{Reference}}
\startdata
$s_{CH}$&4220-4280&4280-4320&-&\citet{BS93}\\
$m_{CH}$&4080-4130&4270-4320&4420-4470&\citet{S96}\\
$CH(G)$&4230-4260&4270-4320&4390-4420&\citet{L99}\\
$S(CH)$&4050-4100&4280-4320&4330-4350&\citet{MSB08}\\
$S(4243)$&-&4290-4318&4314-4322&\citet{B90}\\
$S_{2}(CH)$&4212-4242&4297-4317&4330-4375&\citet{MSB08b}\\
\enddata
\end{deluxetable}

\end{document}